\documentclass[iop]{emulateapj}

\usepackage{graphicx}
\usepackage{hyperref}
\usepackage{amssymb,amsmath}

\def\lea{\mathrel{<\kern-1.0em\lower0.9ex\hbox{$\sim$}}}
\def\gea{\mathrel{>\kern-1.0em\lower0.9ex\hbox{$\sim$}}}
\newcommand{\lta}{{\>\rlap{\raise2pt\hbox{$<$}}\lower3pt\hbox{$\sim$}\>}}
\newcommand{\gta}{{\>\rlap{\raise2pt\hbox{$>$}}\lower3pt\hbox{$\sim$}\>}}

\begin{document}
\title{Hubble Space Telescope detection of the millisecond pulsar J2124$-$3358 and its far-ultraviolet bow shock nebula}


\author{B.\ Rangelov\altaffilmark{1}, G.\ G.\ Pavlov\altaffilmark{2}, O.\ Kargaltsev\altaffilmark{1},  A.\ Reisenegger\altaffilmark{3}, S.\ Guillot\altaffilmark{3}, M.\ van Kerkwijk\altaffilmark{4}, and C.\ Reyes\altaffilmark{3}}

\altaffiltext{1}{Department of Physics, The George Washington University, 725 21st St, NW, Washington, DC 20052}
\altaffiltext{2}{Pennsylvania State University, 525 Davey Lab., University Park, PA 16802}
\altaffiltext{3}{Instituto de Astrofi\'{i}sica, Pontificia Universidad Cat\'{o}lica de Chile, Av.\ Vicu\~na Mackenna 4860, Macul, Santiago, Chile}
\altaffiltext{4}{Department of Astronomy and Astrophysics, University of Toronto, 50 St. George Street, Toronto, ON M5S 3H4, Canada}
\email{blagoy.rangelov@gmail.com}

\slugcomment{The Astrophysical Journal, in press}
\shorttitle{Optical-UV emission from pulsar J2124$-$3358}
\shortauthors{Rangelov et al. 2017}

\begin{abstract}

We observed a nearby millisecond pulsar J2124--3358 with the {\sl Hubble Space Telescope} in broad far-UV (FUV) and optical filters. The pulsar is detected in both bands with fluxes  $F(1250$-$2000\,{\rm \AA})= (2.5\pm 0.3)\times10^{-16}$\,erg\,s$^{-1}$\,cm$^{-2}$ and $F(3800-6000\,{\rm \AA})=(6.4\pm0.4)\times10^{-17}$\,erg\,s$^{-1}$\,cm$^{-2}$, which correspond to luminosities of $\approx 5.8\times 10^{27}$ and $1.4\times 10^{27}$ erg s$^{-1}$, for $d=410$ pc and $E(B-V)=0.03$. The optical-FUV spectrum can be described by a power-law  model, $f_\nu\propto \nu^{\alpha}$, with slope $\alpha=0.18$--0.48 for a conservative range of color excess, $E(B-V)=0.01$--0.08. Since a spectral flux rising with frequency is unusual for pulsar magnetospheric emission in this frequency range, it is possible that the spectrum is predominantly magnetospheric (power law with $\alpha <0$) in the optical while it is dominated by thermal emission from the neutron star surface in the FUV. For a neutron star radius of 12 km, the surface temperature would be between $0.5\times 10^5$ and $2.1\times 10^5$ K, for $\alpha$ ranging from $-1$ to 0, $E(B-V)=0.01$--0.08, and $d=340$--500 pc. In addition to the pulsar, the FUV images reveal extended emission spatially coincident with the known H$\alpha$ bow shock, making PSR J2124--3358 the second pulsar (after PSR J0437$-$4715) with a bow shock detected in FUV.
 
\end{abstract}

\keywords{pulsars: individual (PSR J2124$-$3358) --- shock waves --- ISM: jets and outflows --- ultraviolet: ISM, stars --- X-rays: individual (PSR J2124$-$3358)}

\section{Introduction}

Although many rotation-powered pulsars have been detected in the radio, X-rays and $\gamma$-rays, only about a dozen of them have been detected in the UV-optical-IR (UVOIR) range (see \citealt{2011AdSpR..47.1281M} for a review). Their optical spectra can be described by a power-law (PL) model, $f_{\nu}\propto\nu^{\alpha}$, with slopes $-1\lesssim \alpha\lesssim 0$. The PL components of their X-ray spectra show a faster decrease with frequency, implying spectral break(s) between the optical and X-rays. This  non-thermal  emission is thought to be produced by relativistic electrons/positrons in the pulsar magnetosphere. In addition to the PL optical emission, several middle-aged (a few hundred kyr old) pulsars (e.g., PSR\,B0656+14 and Geminga) exhibit thermal (Rayleigh-Jeans) spectra, $f_\nu\propto \nu^2$, in the far-UV (FUV), originating from the surface of cooling neutron stars (NSs),  with brightness temperatures of $\sim (2$--$8)\times 10^5$ K, usually somewhat lower than those inferred from thermal X-ray components \citep{2007Ap&SS.308..287K}.

As the temperature of a passively cooling NS sharply decreases at ages beyond about 1 Myr \citep{2004ARA&A..42..169Y}, surfaces of old pulsars were expected to be very cold. However, our {\sl Hubble Space Telescope} ({\sl HST}) observations of the 7 Gyr old millisecond (recycled) pulsar J0437--4715 (J0437 hereafter), the only millisecond pulsar firmly detected in the UVOIR range\footnote{Far-UV emission from the double pulsar J0737--3039, detected by \citet{2014ApJ...783L..22D}, likely comes from the millisecond pulsar J0737--3039A, but additional observations are needed to prove it.},  have shown a thermal FUV spectrum emitted from the bulk NS surface with a temperature of about $2\times 10^5$ K \citep{2004ApJ...602..327K,2012ApJ...746....6D}. This result suggests that some heating mechanism(s) operate throughout the life of NSs. To understand the nature of such mechanisms, we initiated an {\sl HST} program to observe old pulsars in the FUV and optical bands. In this paper we report first results from this program, obtained from observations of PSR J2124--3358 (J2124 hereafter).
 
J2124 is a solitary $4.93$\,ms pulsar with a spin-down energy loss rate $\dot{E}=6.8\times10^{33}$\,erg\,s$^{-1}$ and a characteristic age of 3.8\,Gyr \citep{2016MNRAS.455.1751R}. It was discovered during the Parkes 436\,MHz survey of the southern sky \citep{1997ApJ...481..386B}. J2124 has a parallax distance $d=410^{+90}_{-70}$\,pc and an accurately measured proper motion,  $\mu_\alpha \cos\delta = -14.14 \pm 0.04$\,mas\,yr$^{-1}$, $\mu_\delta = -50.08\pm0.09$\,mas\,yr$^{-1}$\citep{2016MNRAS.455.1751R}, corresponding to the transverse velocity $V_\perp=101.2\pm$0.8\,km\,s$^{-1}$ (at $d=410$\,pc). 

X-ray pulsations from J2124 were found by \citet{1999A&A...341..803B} in {\sl ROSAT} observations. {\sl XMM-Newton} observations of J2124 have shown that a two-component  model is required to fit its X-ray spectrum \citep{2006ApJ...638..951Z}. For example, a PL + H-atmosphere (polar cap)  model with  $\Gamma\equiv1-\alpha=2.1\pm0.7$, $T_{\rm pc}=(1.3\pm0.1)\times10^6$\,K, and $R_{\rm pc}=(0.74\pm0.09) d_{410}$\,km provides a reasonable description of the observed spectrum. Alternatively, a two-temperature H-atmosphere model (a hot polar cap ``core'' surrounded by a colder ``rim''), with $T_{\rm core} = (2.2\pm 0.5)\times10^6$\,K, $R_{\rm core}=(0.17\pm 0.4) d_{410}$ km, $T_{\rm rim}=(0.5\pm 0.1)\times10^6$\,K, $R_{\rm rim}=(2.9\pm 1.1) d_{410}$ km, describes the spectrum equally well (see also \citealt{2008ApJ...689..407B}).

J2124 was also detected by  {\sl Fermi} (see the 2PC LAT catalog; \citealt{2013ApJS..208...17A}). Its $\gamma$-ray spectrum can be characterized by a cut-off PL model with $\Gamma= 0.78\pm0.13$ and cut-off energy $E_{\rm cut} = 1.63\pm0.19$\,GeV.  The  corresponding 0.1--100 GeV energy flux is $F_\gamma = (3.68\pm0.16)\times10^{-11}$\,erg\,s$^{-1}$\,cm$^{-2}$. The best-fit $\Gamma$ is rather small for a  pulsar (only 14\% of pulsars in the  2PC LAT catalog have similar or smaller $\Gamma$ values).

In the optical, J2124 was observed by the ESO Very Large Telescope (VLT) in 2001 using the FOcal Reducer and Spectrograph 1 (FORS1) camera.  The pulsar was not detected, with limits of $U\gtrsim26$, $B\gtrsim27.7$, and $V\gtrsim27.8$ \citep{2004AdSpR..33..616M}.

\begin{figure*}
\includegraphics[scale=0.32]{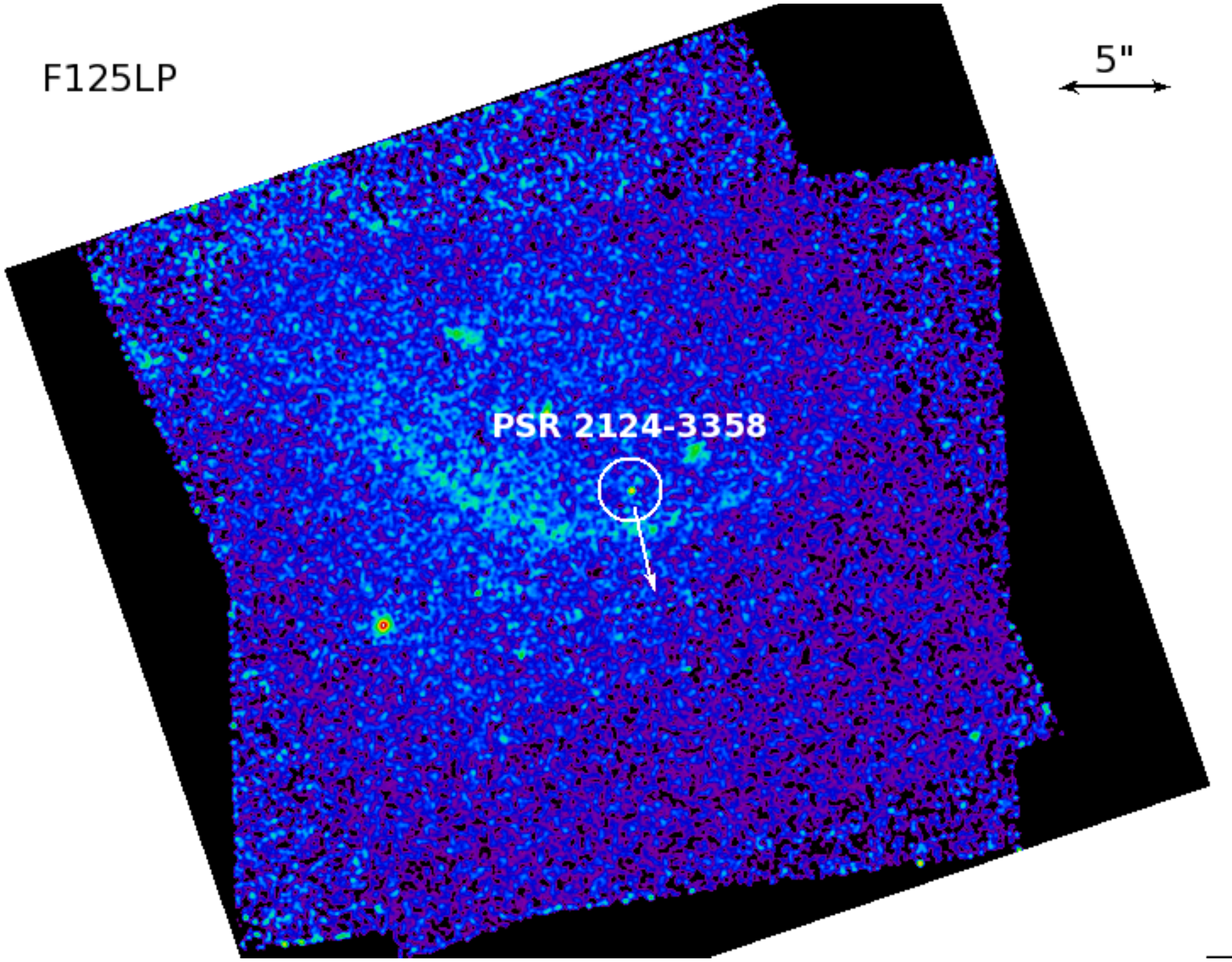}
\includegraphics[scale=0.32]{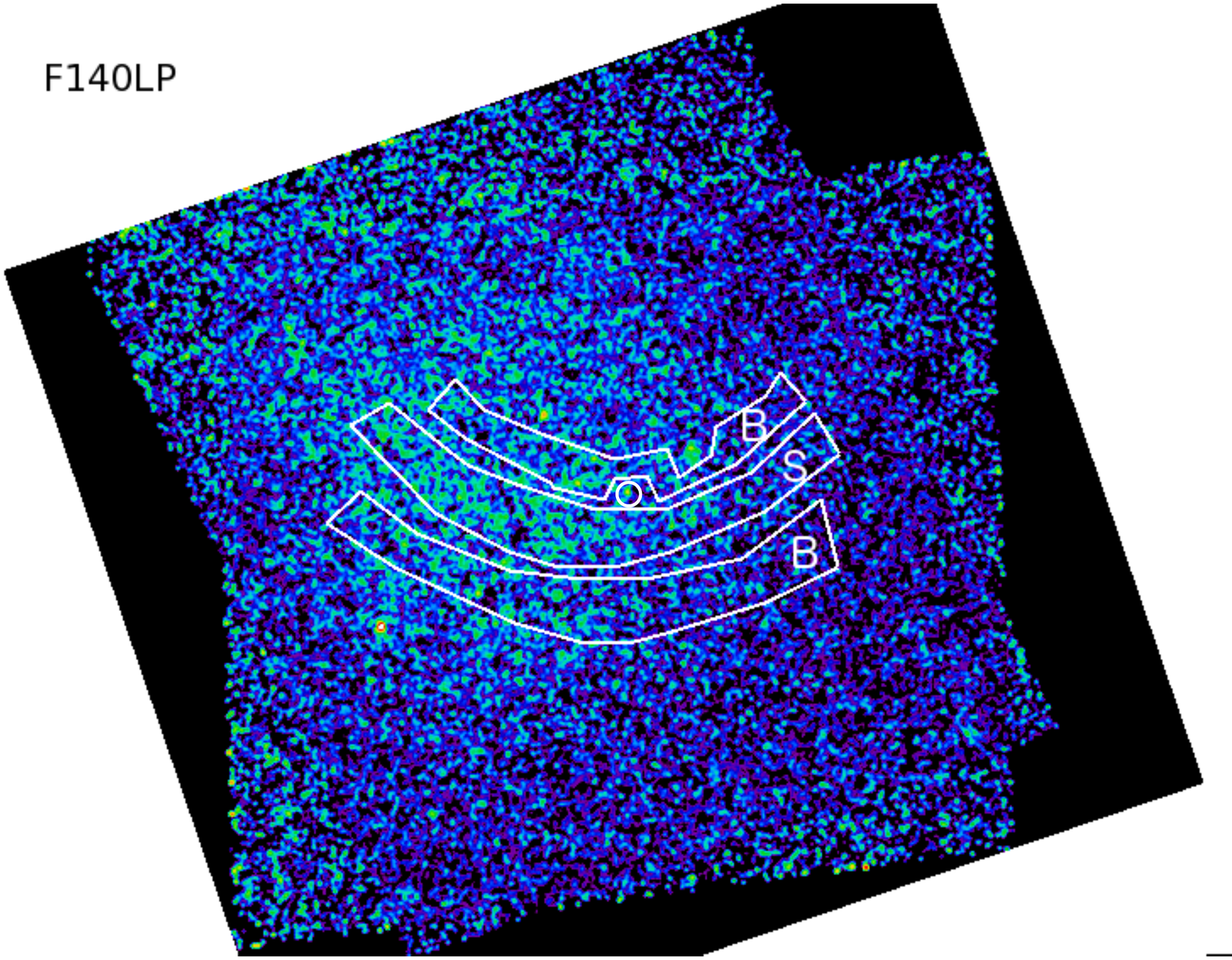}\\
\includegraphics[scale=0.32]{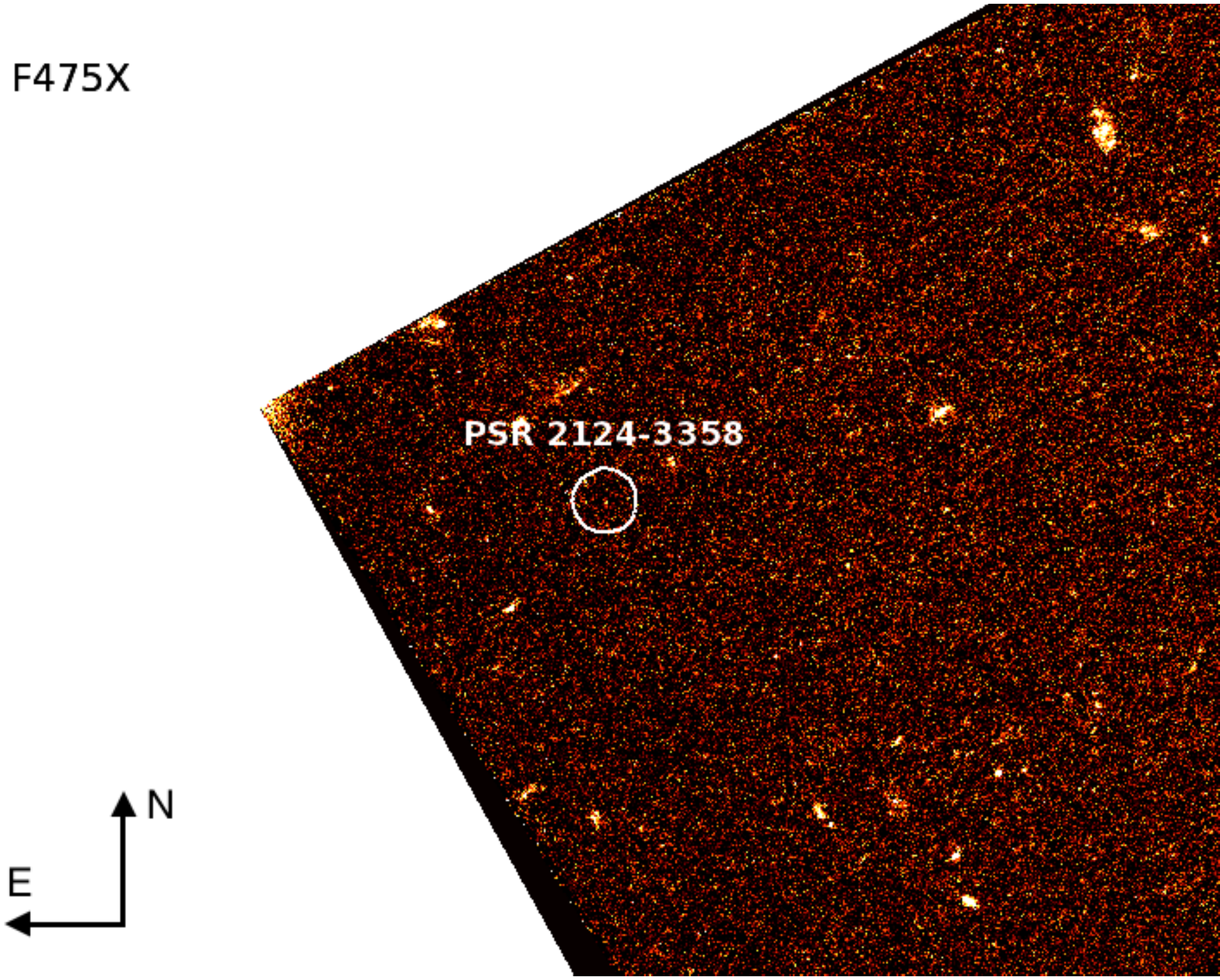}
\includegraphics[scale=0.32]{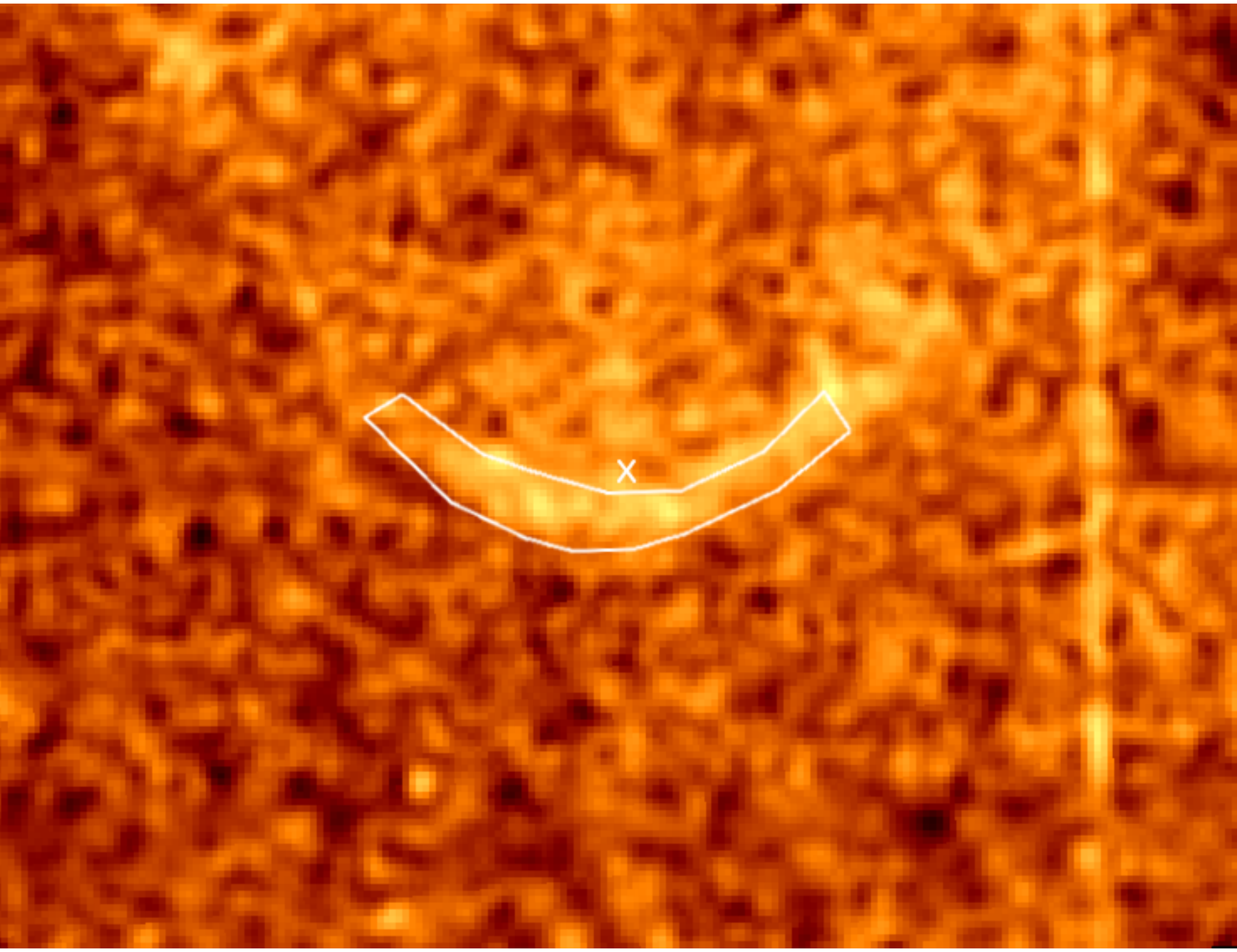}
\caption{
Smoothed {\sl HST}  F125LP (\emph{top left}) and F140LP (\emph{top right}) FUV images of the J2124 field (Gaussian smoothing with $r=0\farcs2$ kernel). J2124 position is marked by small white circle. The bow shock (middle region, labeled as ``S'') and background (two outer  regions, labeled as ``B'') extraction areas are shown in white in the top-right panel, while the pulsar motion is shown with a white arrow in the top-left panel. The SOAR H$\alpha$ (\emph{bottom right}) image (obtained from the SOAR archive and described in detail by \citealt{2014ApJ...784..154B}),  smoothed with $r=1''$  kernel, shows the bow shock emission (J2124 position is marked by the white cross). The bow shock extraction region from the top-right panel is overploted. The bow shock is not visible in the F475X (\emph{bottom left}) image.  All images are to the same scale. North and east are shown in the equatorial coordinate system.
}
\label{hst}
\end{figure*}

\begin{figure}
\includegraphics[scale=0.31]{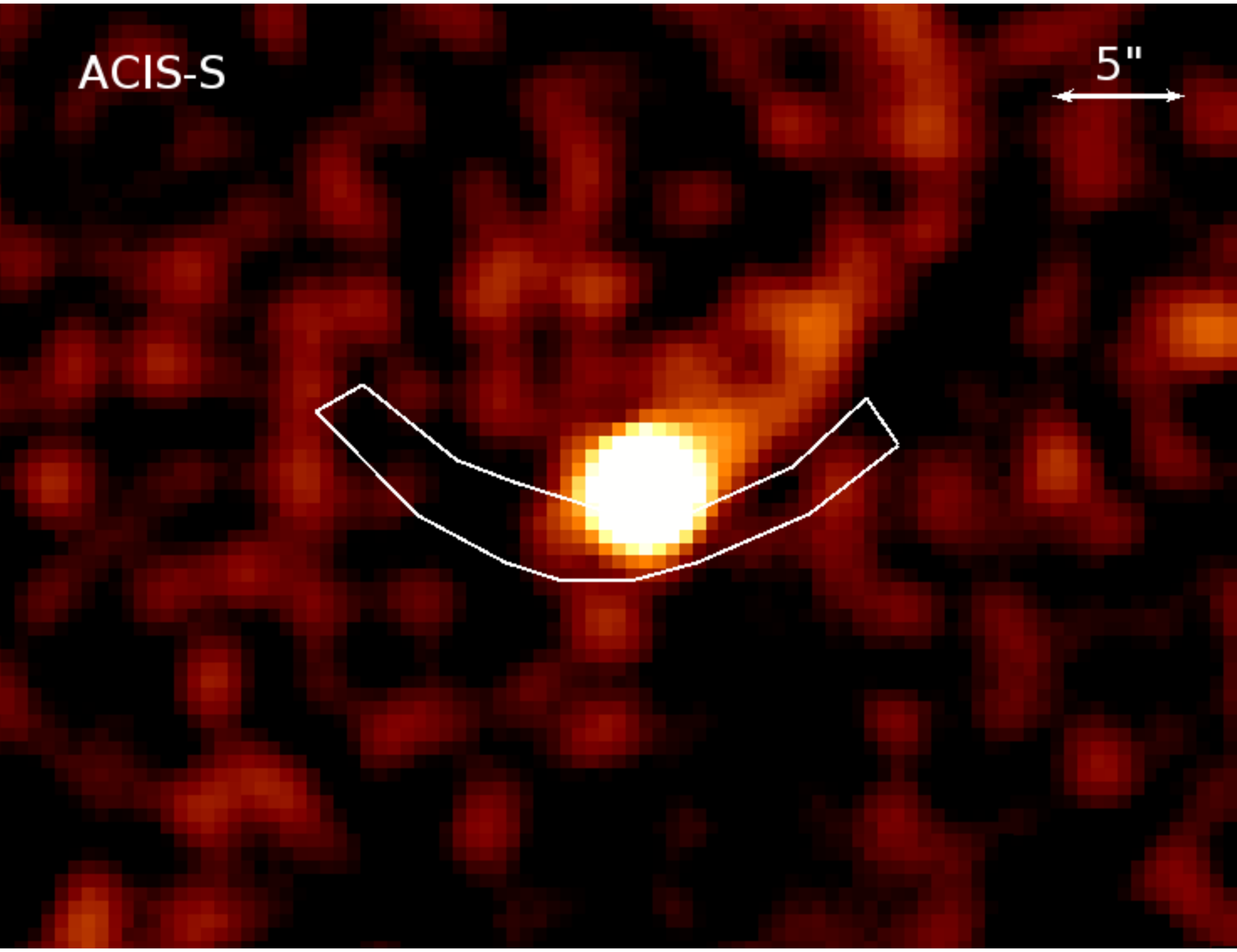}
\caption{{\sl Chandra} ACIS-S smoothed (gaussian kernel $r=2''$) image of J2124. The FUV bow shock extraction region (labeled as `S' in Figure \ref{hst}) is shown for comparison.}
\label{chandra}
\end{figure}

J2124 is among the 9 pulsars (including J0437) around which H$\alpha$ bow-shock nebulae have been detected (\citealt{2014ApJ...784..154B,2002ApJ...580L.137G}; see the H$\alpha$ image in the bottom-right panel of Figure \ref{hst}). H$\alpha$ bow shocks are created by pulsars moving through the interstellar medium (ISM) at a speed exceeding the ISM sound speed, if there are enough neutral H atoms ahead of the pulsar. They are expected to be accompanied by cometary X-ray pulsar wind nebulae, which are produced by synchrotron radiation of the shocked pulsar wind confined by the ram pressure of oncoming medium \citep{2008AIPC..983..171K}. A {\sl Chandra} ACIS observation of J2124 revealed a puzzling faint X-ray nebula which looks like a one-sided, elongated structure projected  within the interior of the  H$\alpha$ bow shock \citep{2005AAS...20718313C,2006A&A...448L..13H}. This X-ray emission extends northwest of the pulsar by $\sim0\farcm5$ (see the X-ray image in Figure \ref{chandra}), and its spectrum fits an absorbed PL with photon index $\Gamma=2.2\pm0.4$ and luminosity $L_{0.5-10\,{\rm keV}}\approx 2.4\times10^{29} d_{410}^2$\,erg\,s$^{-1}$.

In this paper we present an analysis of our {\sl HST} observations of J2124 in optical and FUV bands (Section 2), resulting in detection of the pulsar in both bands (Section 2.2.1), and discovery of  FUV emission from the bow shock (Section 2.2.2).  We discuss the implications of our findings in Section 3.


\section{Observations and Results}

\begin{figure}
\includegraphics[scale=0.6]{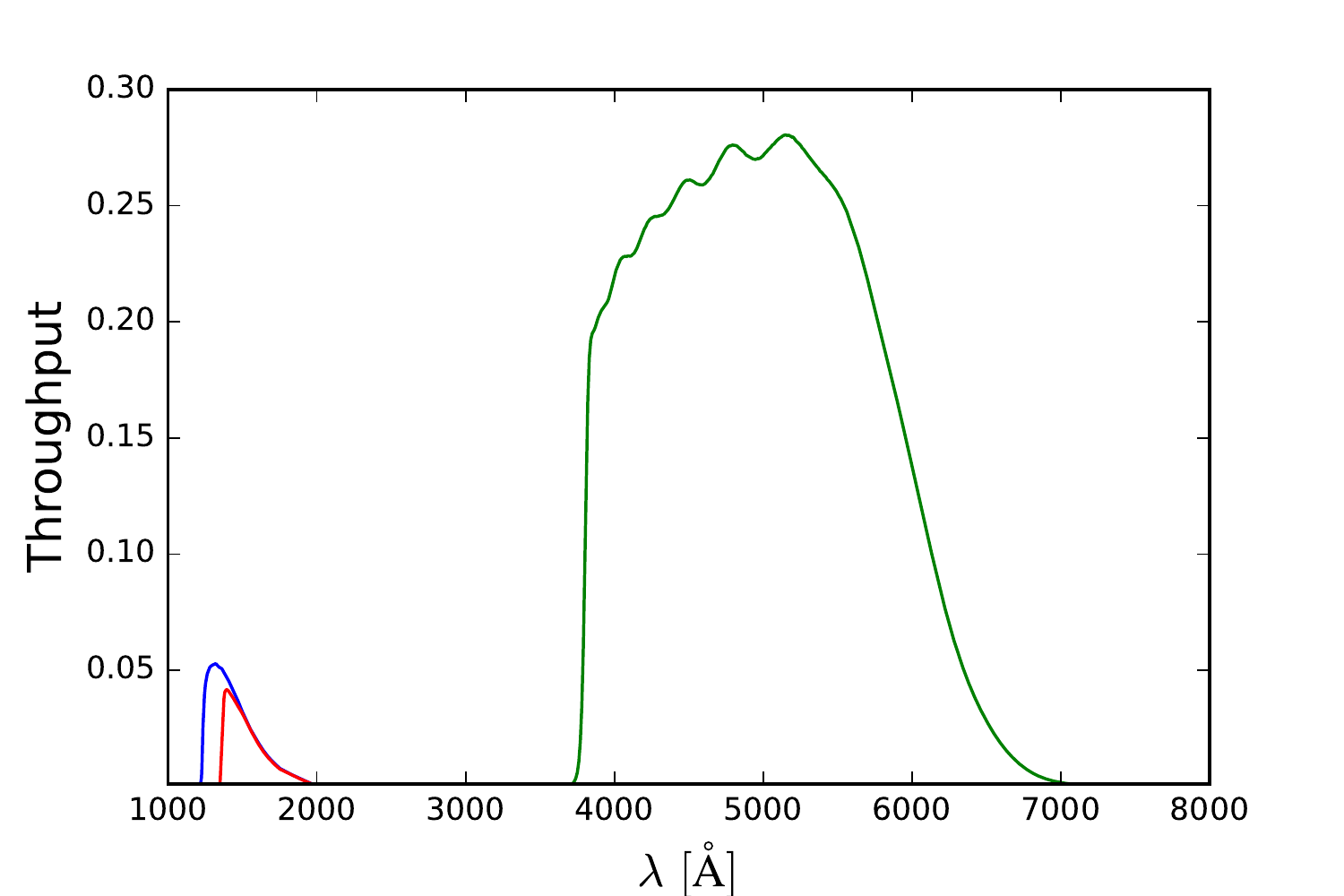}
\caption{
Throughputs of the F125LP, F140LP, and F475X filters (blue, red, and green lines, respectively).
}
\label{throughputs}
\end{figure}

The \emph{HST} observations of J2124 (program 13783, PI Pavlov; see Table~\ref{hstobs}) were carried out in three visits that occurred between 2014 November 15 and 2015 August 12. The data were taken with the Solar Blind Channel (SBC) detector of the Advanced Camera for Surveys (ACS; $34\farcs6\times30\farcs8$ field of view, $0\farcs034\times0\farcs030$ pixel scale) using long-pass filters F125LP and F140LP, and  with the Wide Field Camera 3 (WFC3) Ultraviolet-Visible channel (UVIS; $162''\times162''$ field of view, 0\farcs04 pixel scale) in the very broad F475X (wide $B$) filter. The F125LP, F140LP and F475X filter pivot wavelengths are 1438, 1528, and 4939\,\AA, respectively. The throughputs of these filters are shown in Figure~\ref{throughputs}. The pulsar was placed close to the center of the SBC field-of-view for ACS observations and near the corner ($\approx 11''$ away from the  edges) of the UVIS2 chip in WFC3 observation. The chosen UVIS placement ensures that the pulsar is close to the readout, which minimizes the charge transfer efficiency (CTE) losses\footnote{See Section 6.9 of the WFC3 Instrument Handbook; \url{http://www.stsci.edu/hst/wfc3/documents/handbooks/currentIHB/c06_uvis10.html}.}. The FUV data were acquired during four orbits. The F125LP images were taken in the Earth  shadow, where the geocoronal FUV background is greatly reduced.  The F140LP filter,  which cuts off geocoronal emission shortward of 1400\,\AA, was used outside the shadow parts of the orbits. The UVIS/F475X exposure was  taken during a single orbit. The data were downloaded from the Mikulski Archive for Space Telescopes (MAST\footnote{See \url{http://archive.stsci.edu/}.}). Each MAST image is flat-field corrected using the PyRAF {\tt Multidrizzle} task, which produces co-aligned images,  corrected for geometrical distortion. We perform aperture photometry using the \texttt{PHOT} and \texttt{POLYPHOT} tasks in IRAF\footnote{IRAF is distributed by the National Optical Astronomy Observatories, which are operated by the Association of Universities for Research in Astronomy, Inc., under cooperative agreement with the National Science Foundation.}.

\begin{deluxetable}{cllc}
\tablecaption{\emph{HST} observations of the J2124.\label{hstobs}}
\tablehead{
\colhead{Date} & \colhead{Instrument} & \colhead{Filter} & \colhead{Exposure (s)}}
\startdata
2014-11-15 & WFC3/UVIS & F475X & 2532 \\
2015-07-14 & ACS/SBC & F125LP & 3688 \\
2015-07-14 & ACS/SBC & F140LP & 1218 \\
2015-08-12 & ACS/SBC & F125LP & 3688 \\
2015-08-12 & ACS/SBC & F140LP & 1218 
\enddata
\end{deluxetable}

\subsection{Astrometry}

To identify the pulsar in the \emph{HST} images (see Figure \ref{hst}), we had to check the astrometry of our data. We used stars from the Guide Star Catalog (GSC v.2.3) within the field of view of the WFC3/UVIS to improve the astrometry of the F475X image. Other catalogs with more precise coordinate measurements (e.g., 2MASS, UCAC4) had too few objects in the UVIS field of view. Moreover, most of these objects are bright stars, saturated in the UVIS image, which prevents accurate centroiding and renders them unsuitable for the astrometric correction. Prior to the correction, we found a systematic offset between the GSC v.2.3 and the UVIS stars to be $\approx0\farcs4$ (consistent with the typical {\sl HST} pointing error\footnote{See \url{http://www.stsci.edu/institute/org/telescopes/Reports/Lallo_TIPS_19June08.pdf}}). We then selected  20 stars from the GSC with obvious UVIS counterparts  suitable for accurate centroid determination. As these stars did not have proper motion measurements, we could not adjust their coordinates for the epoch of the {\sl HST} observations. This can explain four outliers (i.e., UVIS stars having $\gtrsim$3 times the mean offset from the catalog position), which we interpret as a proper motion effect. These stars were removed from our final sample of reference stars. We then used the IRAF task \texttt{DAOFIND} to find the detector coordinates of the centroids for the remaining 16  reference stars and supplied them to the IRAF task \texttt{CCMAP}\footnote{See \url{http://stsdas.stsci.edu/cgi-bin/gethelp.cgi?ccmap}.} to calculate new astrometric solution by matching these stars with their GSC counterparts. The refined astrometric solution has standard deviations $\sigma_\alpha=0\farcs17$ and $\sigma_\delta=0\farcs16$ for the sample of 16 stars (at 68\% confidence).


In the SBC image only five sources are seen. Since none of them have counterparts in catalogs, we had to align the F125LP and F140LP images with the UVIS image. Three of the  sources, seen in both  the UVIS and SBC images,  were used in the alignment process.  However, all three sources  appear to be extended  in both images (likely, they are background galaxies). This hinders accurate centroid determination leading to an uncertainty\footnote{Note that \texttt{CCMAP} calculates a ``perfect'' (no error) astrometric solution from three points (stars). The quoted uncertainty, $\approx0\farcs1$, was obtained by selecting different centroiding methods (brightness peak, weighted brightness center, brightest pixel), and selecting the mean and standard deviation as the true center and astrometric solution uncertainty, respectively.} of $\approx0\farcs1$ in aligning the SBC images with  the UVIS image.

In the astrometry-corrected UVIS image we found a faint source located at $\alpha=21$:24:43.841(16) and $\delta=-$33:58:45.01(18). These coordinates are offset by $\Delta\alpha\cos\delta=-0\farcs01\pm0\farcs19$ and $\Delta\delta=-0\farcs18\pm0\farcs18$ from the pulsar's radio coordinates ($\alpha=21$:24:43.841662(24) and $\delta=-$33:58:45.1897(5)) expected at the epoch of the UVIS observation (MJD 56976). Since the offset is within the alignment uncertainty, we conclude that this is a very viable candidate for the pulsar counterpart\footnote{We attempted to improve the astrometry by matching GSC v.2.3 stars with UCAC4 stars  within one degree from the pulsar. Using 370 matches for stars with magnitudes $J\leq12$, we found offsets $0\farcs 012\pm 0\farcs 111$ and $-0\farcs001\pm 0\farcs 076$ along R.A.\ and Decl., respectively. Since these offsets are statistically insignificant, the GSC--UCAC4 matching does not improve the astrometry and does not change our conclusion.}.

\begin{deluxetable*}{lccccc}
\tablecaption{Pulsar and bow shock photometry.\label{photometry}}
\tablehead{
\colhead{Filter} & \colhead{$A_{\rm s}$}
& \colhead{$A_{\rm b}$}
& \colhead{$C_{\rm t}$}
& \colhead{$C_{\rm b}$}
& \colhead{$C_{\rm s}$}\\
\colhead{} & \colhead{arcsec$^2$} & \colhead{arcsec$^2$} & \colhead{cts s$^{-1}$} & \colhead{cts s$^{-1}$} & \colhead{cts s$^{-1}$}
 }
\startdata
\multicolumn{5}{c}{Pulsar}\\
\hline
F125LP & 0.080 & 6.75 & $0.017\pm0.002$ & $0.213\pm0.005$ & $0.014\pm0.002$ \\
F140LP & 0.053 & 6.75 & $0.008\pm0.002$ & $0.198\pm0.009$ & $0.007\pm0.002$ \\
F475X & 0.045 & 0.42 & $0.147\pm0.008$ & $0.061\pm0.005$ & $0.136\pm0.008$ \\
\hline
\multicolumn{5}{c}{Bow shock}\\
\hline
F125LP & 52.1 & 80.5 & $2.96\pm0.02$ & $3.70\pm0.02$ & $0.56\pm0.02$ \\
F140LP & 52.1 & 80.5 & $2.99\pm0.04$ & $3.95\pm0.04$ & $0.43\pm0.04$
\enddata
\end{deluxetable*}

\subsection{Pulsar Photometry}

To find the optimal aperture for the photometry in SBC/F125LP image, we calculated the signal-to-noise ($S/N$) ratio for a set of varying circular  apertures centered on the source. The background was estimated from an annular region with $r_{\rm in}=20$ and $r_{\rm out}=50 $ pixels  (0\farcs64 and 1\farcs6, respectively) centered on the source. We found that the $r=5$ pixels (0\farcs16) aperture provides a maximum $S/N\approx9$. This aperture corresponds to about 64\% of the encircled energy, according to \emph{ACS Instrument Handbook}\footnote{See \url{http://www.stsci.edu/hst/acs/documents/handbooks/current/c05_imaging7.html\#368448}.}. Repeating the process for the F140LP image, we found an optimal aperture ($S/N\approx3$) of $r=4$ pixels  ($0\farcs13$) corresponding to the encircled energy fraction of 58\%. The pulsar detection is less significant due to a factor of 6 smaller exposure time of the F140LP observations and a reduced throughput compared to F125LP. Note that the pulsar photometry was done on individual images to avoid additional broadening of the source caused by the imperfect alignment. 

We repeated this procedure for the UVIS/F475X image, where the background was taken from an annulus with $r_{\rm in}=8 $ and $r_{\rm out}=12$ pixels  (0\farcs32 and 0\farcs48, respectively).  We found  the optimal extraction aperture of  $r=3$ pixels  ($0\farcs12$), which corresponds to the  encircled energy fraction of $\approx78\%$ and provides $S/N\approx19$.  

The net source count rate was calculated as $C_{s}=C_{t} - C_{b} (A_{s}/A_{b})$, where $C_{t}$ is the total count rate in the source region of area $A_{s}$, and $C_{b}$ is the background count rate in the region of area $A_{b}$. Correspondingly, the source count rate uncertainty was calculated as  $\delta C_{s}=\left[ C_{t} + C_{b} (A_{s}/A_{b})^2 \right]^{1/2}t_{\rm exp}^{-1/2}$, where $t_{\rm exp}$ is the exposure time. The net count rates for the optimal apertures are given in Table 2. 

We then corrected the rates for the finite aperture sizes and followed standard photometry procedures to convert count rates to flux densities\footnote{See \url{http://www.stsci.edu/hst/wfc3/phot_zp_lbn}}. The observed mean flux densities for F125LP, F140LP, and F475X are $27\pm4$\,nJy, $25\pm11$\,nJy, and $22\pm2$\,nJy, respectively.  For a plausible reddening $E(B-V)=0.03$ (see Section~2.3) these values correspond to dereddened  flux densities of $35\pm6$\,nJy, $33\pm15$\,nJy, and $24\pm2$\,nJy, for F125LP, F140LP, and F475X, respectively.

\subsection{Extinction}

To interpret the photometry results, we need to know the interstellar extinction toward J2124. The extinction coefficient $A_\nu$ is proportional to the color excess $E(B-V)$, which, on average, is proportional to the hydrogen column density $N_{\rm H}$. \cite{2006ApJ...638..951Z} reported $N_{\rm H}=(1$--$3)\times10^{20}$\,cm$^{-2}$ toward J2124 based on fits to the X-ray data. The dispersion measure, ${\rm DM}=4.6$ pc cm$^{-3}$,   leads to a similar $N_{\rm H}=1.4\times10^{20}$ cm$^{-2}$ (for an assumed 10\% degree of ISM ionization). The relation $N_{\rm H} = 2.2\times10^{21} R_V E(B-V)$ cm$^{-2}$   \citep{1975ApJ...198...95G,2009MNRAS.400.2050G}, with the commonly assumed $R_V=3.1$,  corresponds to  $E(B-V)=0.015$--$0.045$. In our analysis below, we explore an even broader range of color excess, $E(B-V)=0.01$--0.08, to account for the additional uncertainty introduced by the $N_{\rm H}$-to-$E(B-V)$ conversion. The upper bound on color excess, $E(B-V) = 0.08$, in the direction to J2124 ($l=10\fdg9$, $b=-45\fdg4$) is based on the Galactic extinction maps by \citet{1998ApJ...500..525S}. To perform  de-reddening, we use the extinction curves adopted from \citet{2003ApJ...588..871C}.

\subsection{Pulsar Spectral Fits}

\begin{figure}
\includegraphics[scale=0.6]{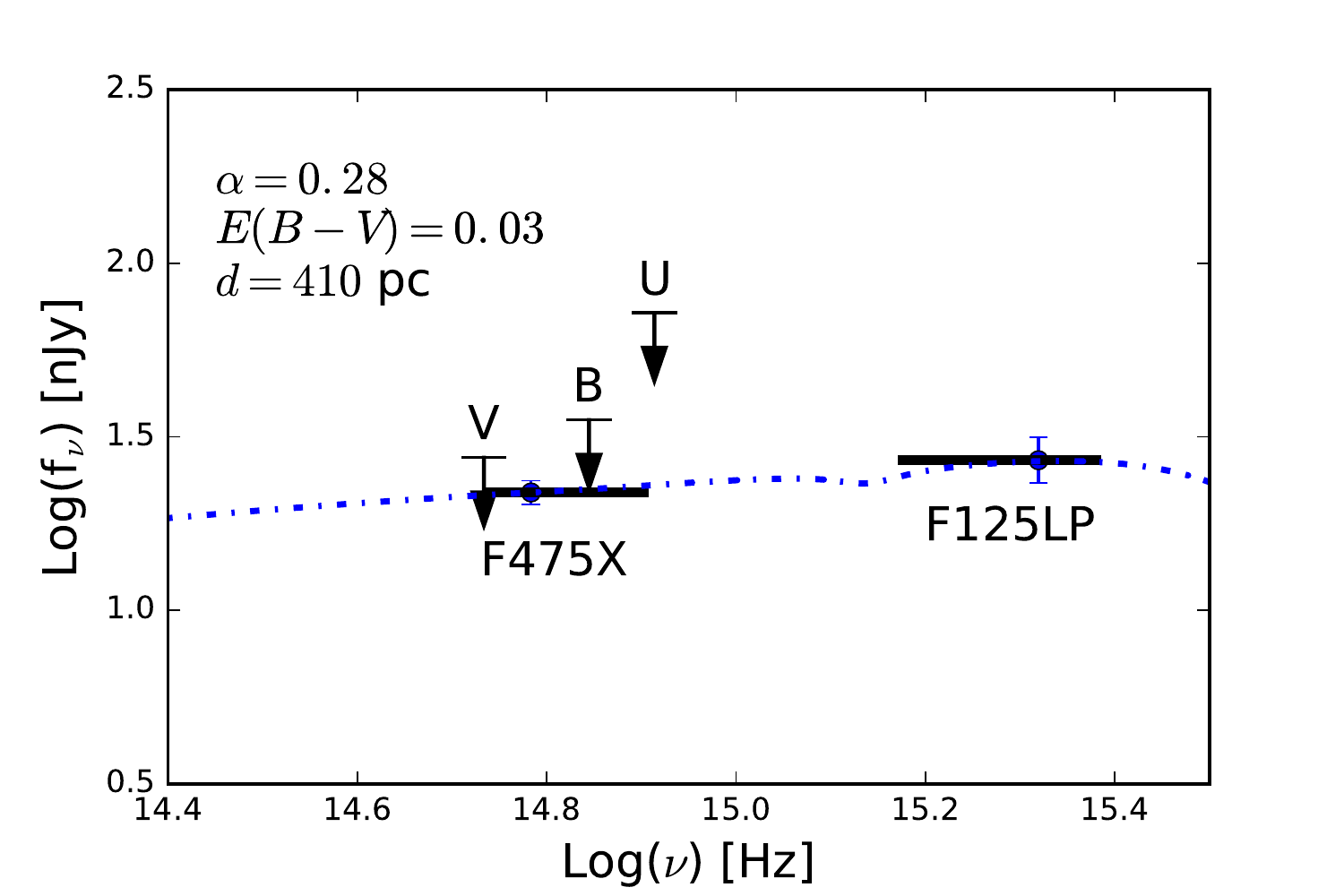}
\caption{Example of a single PL fit to the {\sl HST} photometry results for the J2124 pulsar (the measurement in F140LP is not shown because its band is within the F125LP band). The blue dash-dot line shows the best-fit model spectrum at fixed $d=410$ pc, reddened with ${E(B-V)}=0.03$. The downward arrows show the VLT limits reported by \citet{2004AdSpR..33..616M}. The errors for the F475X data point are comparable to the size of the dot. The black horizontal lines show the filter widths at half maximum.}
\label{pl-only}
\end{figure}

\begin{figure}
\includegraphics[scale=0.6]{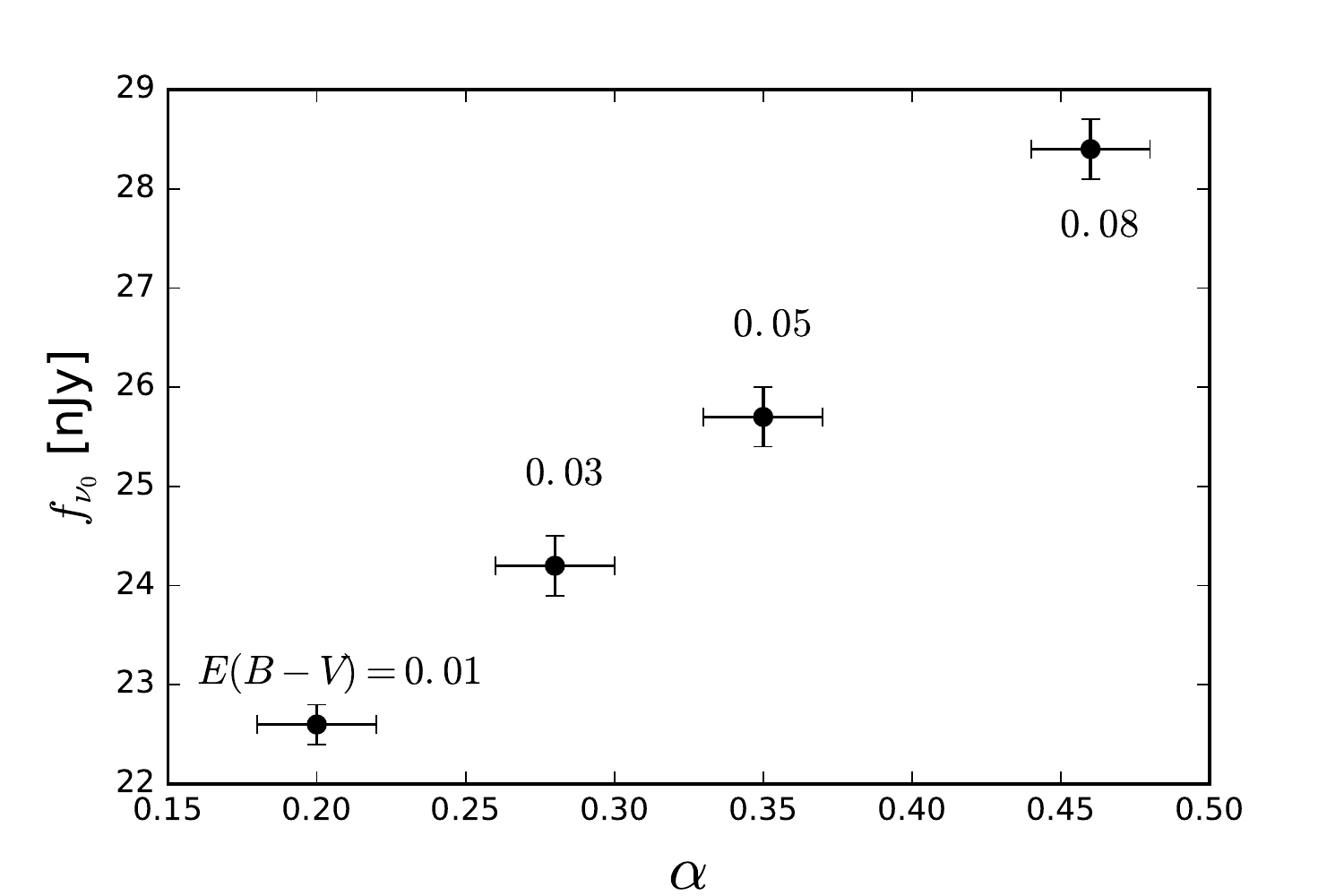}
\caption{Normalization $f_{\nu_0}$ at $\nu_0=6.07\times 10^{14}$ Hz as a function of slope $\alpha$ for the single PL  model fits at different values of 
$E(B-V)$.}
\label{alpha-n}
\end{figure}

The photometric data in only 3 spectral bands (of which one, F140LP, is within the other, F125LP) can be fitted with many spectral models. We first fit the data with an absorbed PL model, $f_\nu=f_{\nu_0}(\nu/\nu_0)^{\alpha}\times10^{-0.4 A_\nu}$, which is often used to describe magnetospheric emission \citep{1997ApJ...489L..75P}. We fit  the PL slope $\alpha$ and normalization $f_{\nu_0}$ at $\nu_0=6.07\times 10^{14}$ Hz (corresponding to the pivot wavelength of the F475X filter) for 4 fixed values of color excess: $E(B-V)=[0.01, 0.03, 0.05, 0.08]$. An example of such a fit is shown in Figure~\ref{pl-only}, and the PL parameters for the four  $E(B-V)$ values are given in  Figure~\ref{alpha-n}. The slope and normalization values on the $E(B-V)$ grid are $\alpha=[0.20\pm0.02, 0.28\pm0.02, 0.35\pm0.02, 0.46\pm0.02]$ and $f_{\nu_0}= [22.6\pm0.2, 24.1\pm0.3, 25.7\pm0.3, 28.4\pm0.3]$ nJy (at $6.07\times10^{14}$\,Hz), respectively. Their dependences on color excess can be approximated by linear functions in this $E(B-V)$ range: $\alpha\approx 0.164 + 3.71\,E(B-V)$ and $f_{\nu_0} \approx 21.6 +82.9\,E(B-V)$ nJy. The corresponding de-reddened fluxes in the FUV and optical bands are $F(1250$--$2000\,{\rm \AA})= [2.6\pm 0.3, 2.9\pm 0.3, 3.4\pm 0.4, 4.4\pm 0.5] \times10^{-16}$ and $F(3800$--$6000\,{\rm \AA}) = [6.6\pm 0.4, 7.1\pm 0.4, 7.6\pm 0.4, 8.4\pm 0.5] \times 10^{-17}$\,erg\,s$^{-1}$\,cm$^{-2}$.

We then considered a single thermal (BB) model with a fixed radius $R=12$\,km, a typical radius of a NS as seen by a distant observer. We fit both the BB temperature $T$ and color excess $E(B-V)$ for distances $d=340$--500\, pc. Because the FUV band falls in the Rayleigh-Jeans regime for the obtained temperatures, we can express the final results as  $T=(3.6\pm0.2)\times10^6 d_{410}^2$\,K and $E(B-V)=0.51\pm0.01$. Since this color excess is much larger than the Galactic value along the line-of-sight and the temperature is implausibly high for the entire NS surface (contradicts to the X-ray data), this model can be ruled out.

\begin{figure}
\includegraphics[scale=0.64]{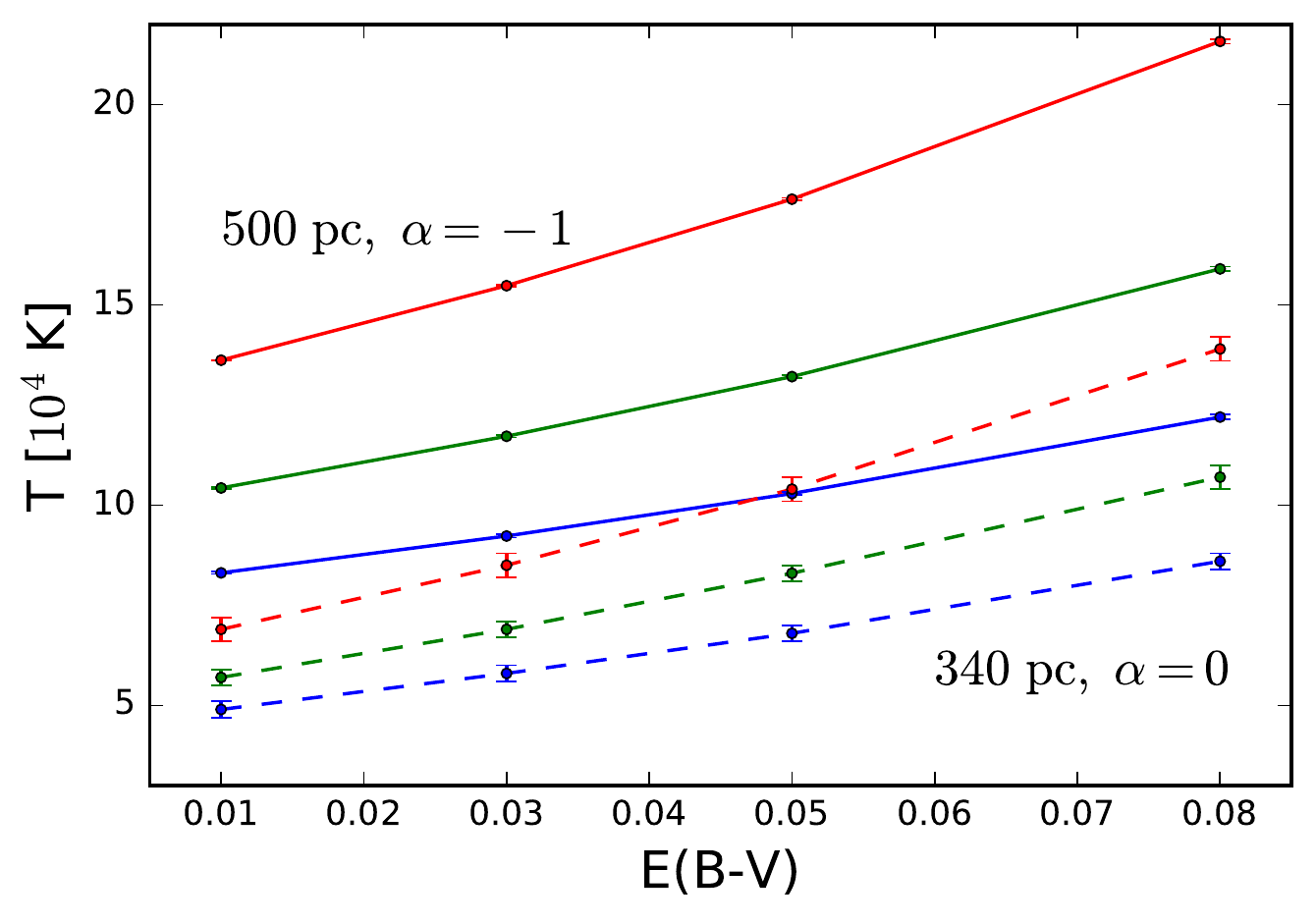}\\
\includegraphics[scale=0.64]{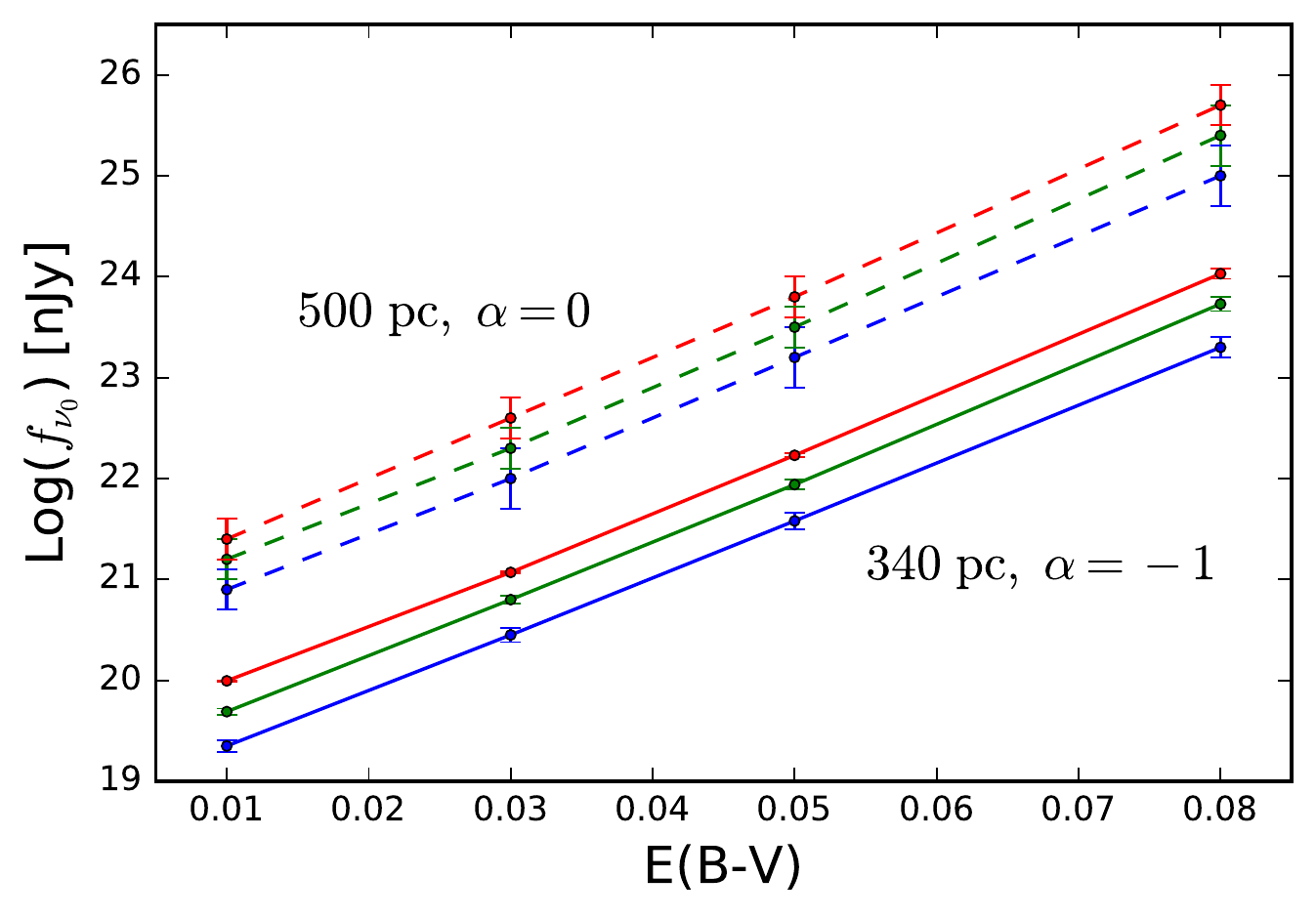}
\caption{BB temperature (upper panel) and PL normalization at $\nu_0=6.07\times 10^{14}$\,Hz (lower panel) in the BB+PL fits  as functions of $E(B-V)$ for different values of  $d$ and $\alpha$. The blue, green, and red solid lines correspond to the models for $d=[340, 410, 500]$\,pc and $\alpha=-1$. The blue, green, and red dashed lines are for $\alpha=0$ and the same values of distance.}
\label{t_ebv}
\end{figure}

\begin{figure*}
\includegraphics[scale=0.6]{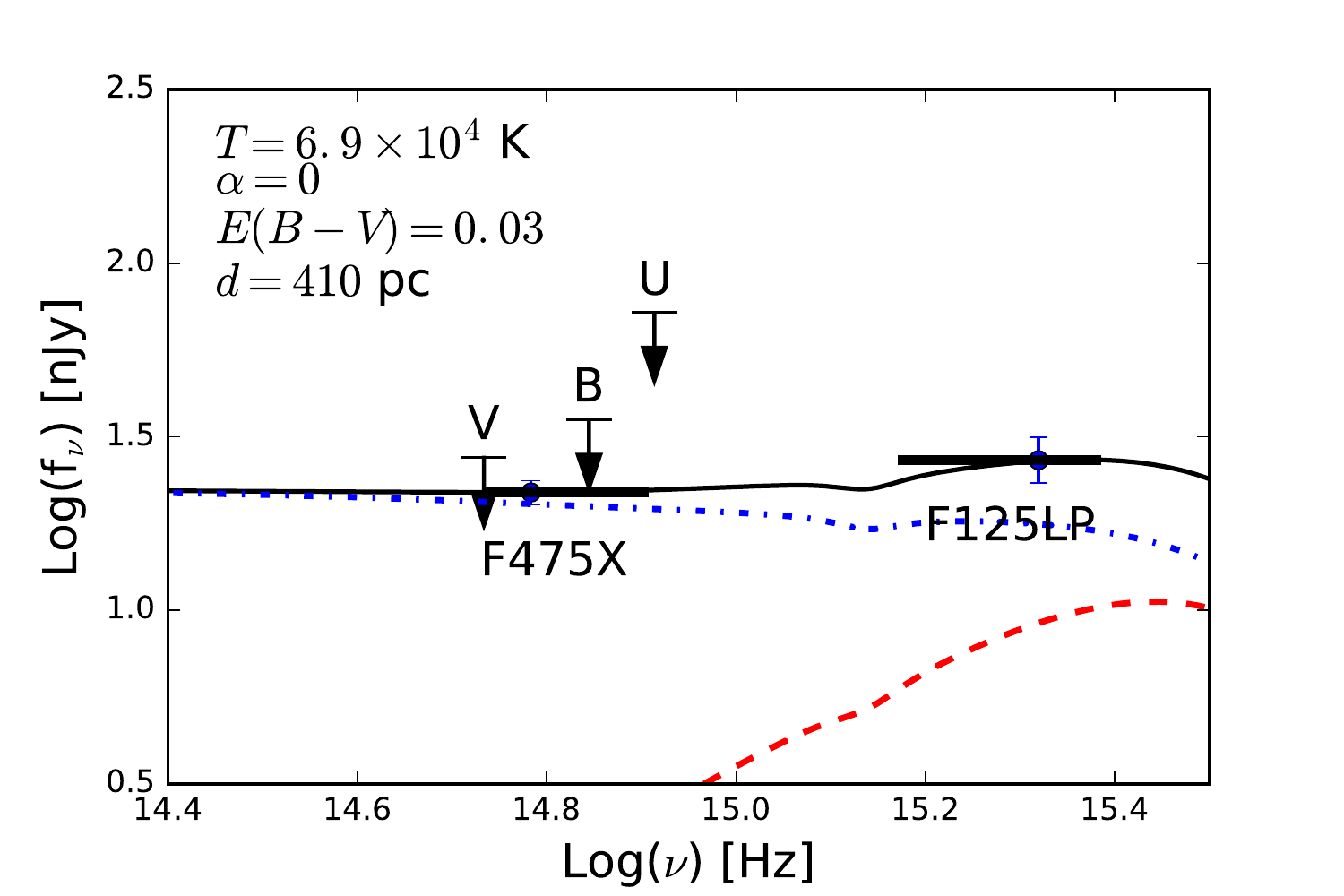}
\includegraphics[scale=0.6]{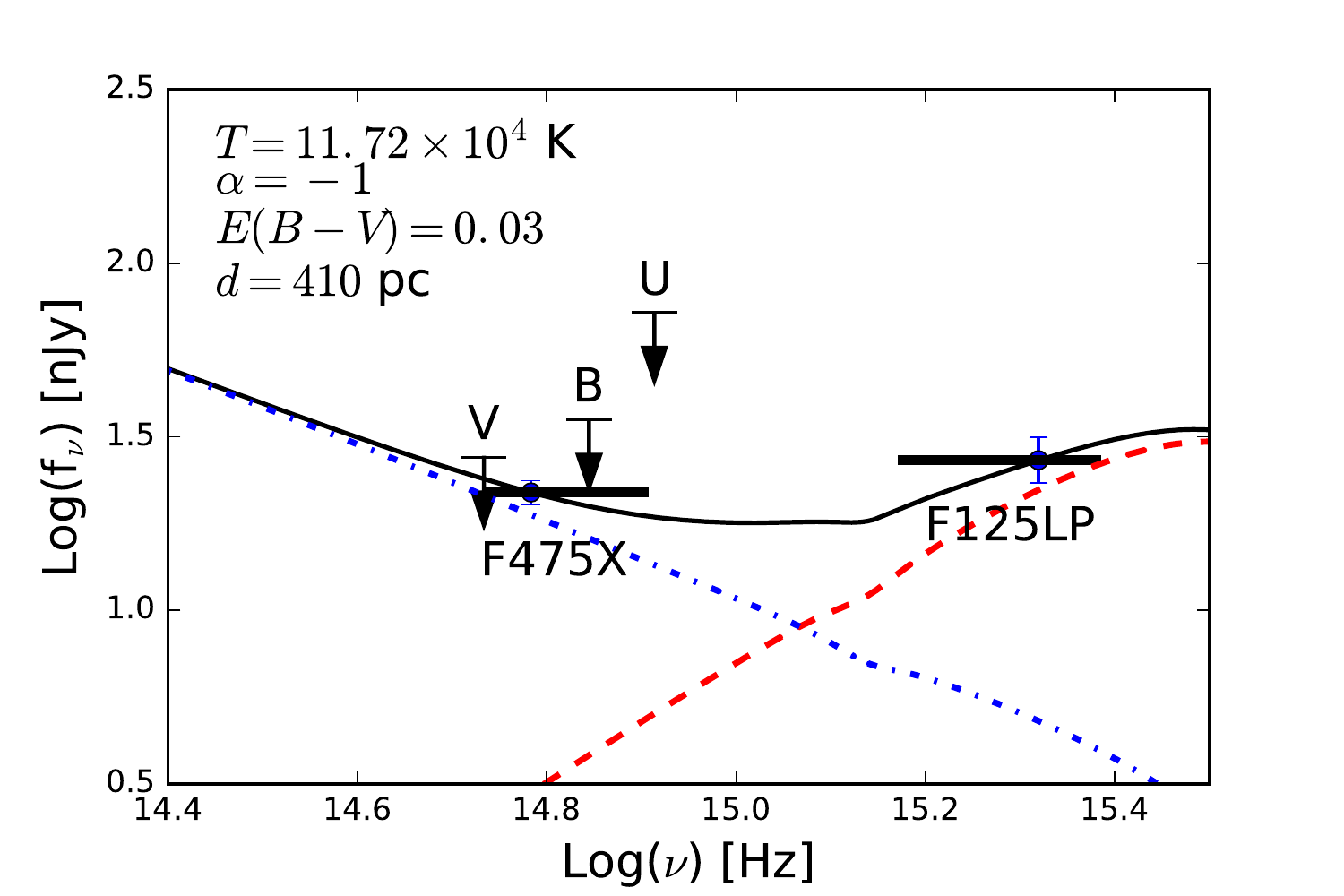}
\caption{
Examples of the BB+PL fit to the {\sl HST} photometry results for the J2124 pulsar, for two values of the PL slope, $\alpha=0$ and $-1$, at $d=410$ pc and $E(B-V)=0.03$. The red dashed lines and blue dash-dot lines show the BB and PL components, respectively, while the black solid lines show their sums. The downward arrows are the VLT limits reported by \citet{2004AdSpR..33..616M}. The black horizontal lines show the filter widths at half maximum.}
\label{temperature}
\end{figure*}

Lastly, we fit a two-component BB+PL model. Given the small number of data points, we fit the BB temperature $T$ and PL normalization $f_{\nu_0}$ on a three-dimensional grid $\alpha=[-1, 0]$, $d=[340, 410, 500]$\,pc, and $E(B-V)=[0.01, 0.03, 0.05, 0.08]$. The resulting BB temperatures and PL normalizations are plotted in Figure~\ref{t_ebv}.  Examples of best-fit model spectra at most plausible $d=410$\,pc, $E(B-V)=0.03$ are shown in Figure~\ref{temperature} for two values of PL slope, $\alpha=0$ and $-1$.

\subsection{Bow shock}

A bow-shaped structure spatially coincident with the known H$\alpha$ bow shock (see \citealt{2002ApJ...580L.137G,2014ApJ...784..154B}) is clearly visible in the F125LP image shown in Figure \ref{hst} (top left).  The FUV  bow shock appears to be asymmetric,  being thicker and brighter east of the pulsar. We measure the apex distance (from the pulsar to the leading edge of the bow shock) to be $\approx2\farcs5$, consistent with the  \citet{2014ApJ...784..154B,2002ApJ...580L.137G} measurements in H$\alpha$. 

To extract the bow shock flux, we define the source and two background regions, ``inside'' and ``outside'' of the bow shock,  shown in Figure \ref{hst} (top right). 
The background regions were selected based on the F125LP image, in which the bow shock was most clearly detected. The regions are close to the bow shock extraction region to mitigate effects from the inhomogeneous detector background while excluding the pulsar. Due to the larger apparent thickness of the east side of the bow shock, more space was left between the bow shock extraction region and the background regions. We take into account the background variations by selecting background regions on both sides of the bow shock:  one ?inside? and one ?outside? the bow shock.
These regions  were used for both F125LP and F140LP flux measurements. A combined background from the two regions was extracted. The net source count rate was calculated similarly to the pulsar, with the area $A_{b}$ being the combined area of the ``outside'' and ``inside'' background regions. The F125LP and F140LP count rates and their uncertainties are given in Table~\ref{photometry}. From the measured count rates we calculated the absorbed fluxes in the F125LP and F140LP wavelength ranges: $F(1250-2000\,{\rm \AA})= (8.1\pm 0.4) \times10^{-15}$\,erg\,s$^{-1}$\,cm$^{-2}$ and $F(1350-2000\,{\rm \AA})= (7.0\pm0.7) \times10^{-15}$\,erg\,s$^{-1}$\,cm$^{-2}$, for a flat $f_\lambda$ spectrum. The unabsorbed luminosity can be estimated as  $L(1250-2000\,{\rm \AA}) \approx 1.9 \times10^{29}$\,erg\,s$^{-1}$, for $E(B-V)=0.03$, and $d=410$\,pc.

\begin{figure}
\includegraphics[scale=0.34, trim=30 30 30 30]{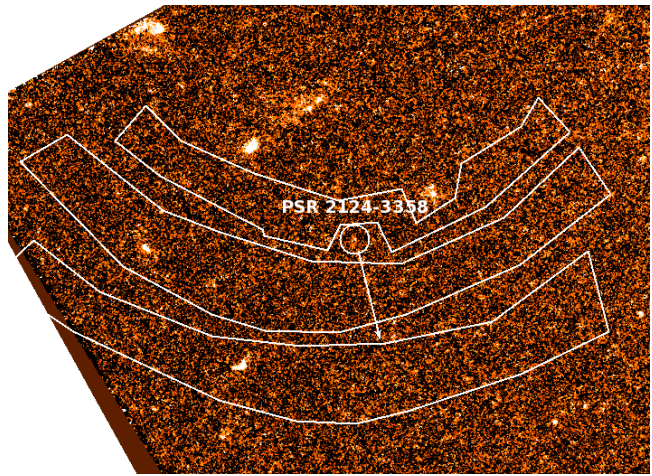}
\caption{
UVIS/F475X image with the bow shock extraction regions (the same as in Figure~\ref{hst}), used for the 
data analysis. The pulsar and the direction of its proper motion are shown with a circle and an arrow, respectively. 
}
\label{f475x}
\end{figure}

The bow shock is not detected in the F475X images (see Figure~\ref{f475x}). Because of the apparent non-uniformities in the background, which significantly exceed the statistical fluctuations, we chose to measure the upper limits by sampling the background counts from ten different regions in the vicinity of the bow shock and calculating the standard deviation from the mean value. The areas of the background regions were equal to that of the bow shock region used for FUV flux measurements. The standard deviation can be considered as  a conservative $1\sigma$ upper limit. The corresponding $3\sigma$ upper limit on the bow shock flux in the area of 52 arcsec$^2$ is $F(3750-6000\,{\rm \AA})\approx 2.9\times10^{-15}$\,erg\,s$^{-1}$\,cm$^{-2}$, for a flat  $f_\lambda$ spectrum and $E(B-V)=0.03$.

Similar to our analysis of the FUV bow shock around J0437 \citep{2016ApJ...831..129R}, we measured the ratio of the count rates in F125LP and F140LP filters for the J2124 bow shock. We obtained a ratio of $C_{125}/C_{140}=1.3\pm0.2$. It is smaller than $C_{125}/C_{140}=1.63\pm0.08$ for the J0437 bow shock, but the associated uncertainty is large.

\section{Discussion}

\subsection{Spectrum of PSR~J2124$-$3358}

The obtained photometric measurements can be used to constrain the optical-FUV spectral energy distribution (SED) for J2124 and to look for thermal emission, similar to that of J0437. 

The slope $\alpha \gtrsim 0.2$ inferred from  the nonthermal (single PL) fit is outside the typical range of $-1\lesssim\alpha\lesssim0$ found for the optical-UV magnetospheric emission seen in other pulsars  \citep{2007Ap&SS.308..287K,2011AdSpR..47.1281M}. However, we cannot rule out the possibility that the FUV-optical emission of J2124 is entirely magnetospheric, with $\alpha >0$, because virtually nothing is known about optical spectra of non-accreting recycled pulsars\footnote{Optical emission from the J0437 pulsar is buried under the much brighter emission of the white dwarf companion \citep{2012ApJ...746....6D}.}.  The  PL fit implies the V-band luminosity $L_{\rm V}\approx 5\times10^{26}d_{410}^2$ erg s$^{-1}$, which corresponds to an optical efficiency\footnote{The observed spin-down energy loss rate, $\dot{E}$, needs to be corrected for the Shklovskii effect \citep{1970SvA....13..562S}.  With the corrected  period derivative, $\dot{P}_{\rm int}=2.06\times10^{-20}$\,s\,s$^{-1}$, we obtain the intrinsic spin-down power $\dot{E}_{\rm int}=(2.4^{+0.6}_{-0.9})\times10^{33}$\,erg\,s$^{-1}$ (for the the moment of inertia of $10^{45}$\,g\,cm$^2$) and the characteristic pulsar age of $11_{-3}^{+6}$\,Gyr, for $d=410^{+90}_{-70}$ pc.} $\eta_V\equiv L_V/\dot E_{\rm int} \sim 2\times 10^{-7}$, similar to  the efficiencies of middle-aged pulsars (see Figure~4 in \citealt{2012A&A...540A..28D}).
 
A thermal interpretation for combined FUV and optical emission (single BB fit) can be ruled out because it requires a color excess much larger  than the  Galactic value along this line-of-sight, as well as an implausibly high temperature, $\sim 3.6\times10^6$\,K, of the entire NS surface, which would result in an X-ray luminosity much larger than observed.
 
Finally, there remains a good possibility that the optical emission is predominantly nonthermal while the FUV emission is largely thermal, similar to what is observed for nearby middle-aged pulsars \citep{2007Ap&SS.308..287K}. The optical efficiency in the PL+BB fit is similar to (slightly lower than) that of the single PL fit, because the thermal component contribution is very small in the optical (see Figure \ref{temperature}). It follows from Figure \ref{t_ebv} that the NS surface temperature of J2124 is likely in the range $T\approx (0.5$--$2.1)\times10^5$\,K for the considered (conservative) range color excess values, $E(B-V)=0.01$--0.08, and distances $d=340$--500 pc. Thus, this interpetation of the J2124 spectrum yields
 the NS surface temperature similar to that of J0437 \citep{2012ApJ...746....6D}. 

\citet{2006ApJ...638..951Z} has shown that two components are required to fit the  X-ray spectrum of J2124. However, various combinations of thermal and nonthermal components were found to be statistically acceptable (e.g., two-component H-atmosphere model, H-atmosphere + PL, BB+PL; see Section 1). Extrapolation of the thermal components, originating from small, hot areas on the NS surface (presumably polar caps), underpredicts the observed FUV fluxes by several orders of magnitude. Therefore, the X-ray-emitting heated regions cannot be responsible for the observed FUV emission from the J2124. The extrapolated best-fit X-ray PL components of the BB+PL and H-atmosphere + PL models overpredict the FUV-optical fluxes by a large margin, but this is not unusual for middle-aged pulsars, which often show a spectral break between the X-ray and optical bands. The extrapolated $\gamma$-ray spectrum is a few orders of magnitude below the optical-FUV fluxes, which means that the optical-FUV and $\gamma$-ray
emission are produced by different mechanisms.

If the two-component (thermal plus magnetospheric) nature of the optical-FUV spectrum of J2124 is confirmed in future observations, J2124 would be the second recycled pulsar, after J0437, with such a high temperature, $T\sim 10^5$ K of the NS surface. It would firmly prove that a heating mechanism operates at least in millisecond pulsars and could distinguish between heating models.

Heating mechanisms that may operate in the interiors of ancient NSs include the dissipation of rotational energy due to interactions between superfluid and normal components of the NS (vortex creep;  \citealt{1984ApJ...276..325A,1989ApJ...346..808S}), release of strain energy stored by the solid crust due to spin-down deformation \citep{1992ApJ...396..135C}, rotochemical heating \citep{1995ApJ...442..749R,2005ApJ...625..291F,2010A&A...521A..77P}, rotation-induced deep crustal heating (a variant of rotochemical heating that operates in the crusts of pulsars; \citealt{2015MNRAS.453L..36G}), and decay or annihilation of dark matter particles, trapped in the NS interiors \citep{2010PhRvD..81l3521D,2010PhRvD..82f3531K}. Among the internal heating mechanisms, the vortex creep, rotochemical heating, and deep crustal heating, all of which are driven by the spin-down of pulsars, are able to account for the high surface temperatures, $\sim 10^5$ K, of millisecond pulsars \citep{2010A&A...522A..16G,2015MNRAS.453L..36G}. Both J2124 and J0437 are much older than the standard NS cooling time, so they are expected to be in a quasi-stationary state in which any internal heating and photon cooling balance each other (e.g., Reisenegger 1995; Fern\'andez \& Reisenegger 2005). Thus the effective temperature should be a function of the spin-down parameters, $\propto\dot\Omega^{1/4}$ for vortex creep, $\propto(\Omega\dot\Omega)^{2/7}$ for rotochemical heating (assuming modified Urca reactions and neglecting superfluid corrections), and $\propto(\Omega\dot\Omega)^{1/4}$ for rotation-induced deep crustal heating, where $\Omega$ is the NS rotation rate and $\dot\Omega$ is its time derivative. The spin-down parameters (corrected for the Shklovskii effect) of J2124 are quite similar to those of J0437, so the ratio of their effective temperatures is expected to be $T_{\rm J2124}/T_{\rm J0437}=0.89, 0.95, 0.96$, respectively, for the three heating mechanisms. Since, for a radius of 12 km (as assumed here), $T_{\rm J0437}\approx 1.8\times 10^5$ K (inferred from the results of \citealt{2012ApJ...746....6D}), the temperature expected for J2124 is only slightly lower, $T_{\rm J2124}\sim (1.6-1.7)\times 10^5$ K, which is consistent with temperatures estimated in the PL+BB scenario. To confirm this scenario and estimate the temperature more accurately, the J2124 pulsar must be observed with several filters in the UVOIR range, and the distance to the pulsar should be measured more precisely.

\begin{figure}
\includegraphics[scale=0.58]{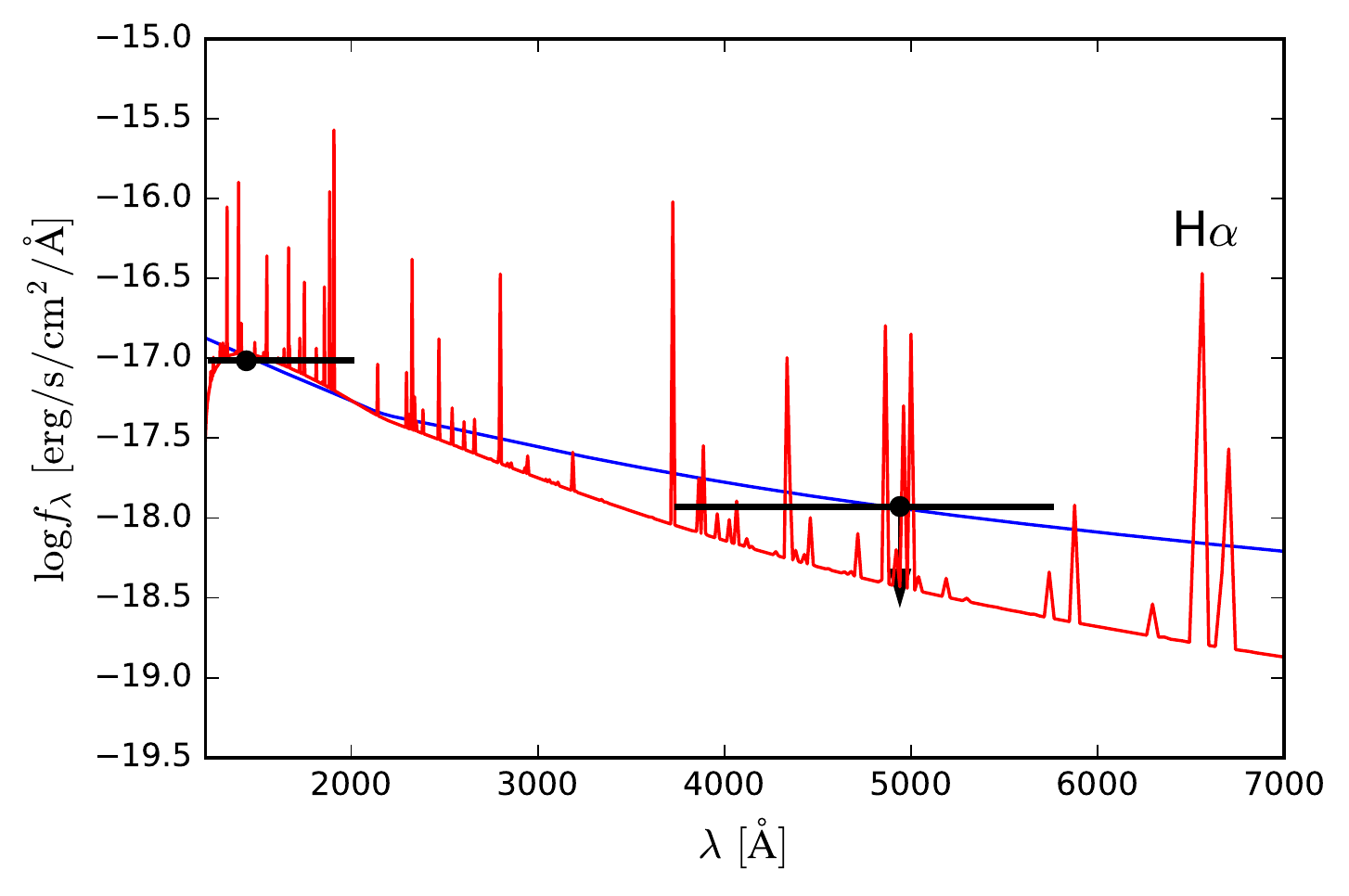}
\caption{The bow shock flux density in FUV (F125LP) and F475X $3\sigma$ upper limit (see Section 2.5 for details) are shown together with two spectral models. The horizontal black lines represent the filter widths at the half-maximum. Absorbed flux densities for the PL ($\alpha=0$) and shocked plasma model (for $v_s=110$\,km\,s$^{-1}$ and $n_0=0.2$\,cm$^{-3}$; see \citealt{2016ApJ...831..129R} for details) are plotted as blue and red lines, respectively.}
\label{limits}
\end{figure}

\subsection{FUV bow shock}

J2124 is the second pulsar with a bow shock detected in FUV. The first FUV bow shock was discovered by \citet{2016ApJ...831..129R} around J0437. These authors tested two spectral models for FUV emission. A PL continuum model, with $-3\lesssim\alpha \lesssim -1$ in the FUV range ($\alpha \gtrsim -1$ if the PL spectrum extends to the optical range), could correspond to synchrotron radiation of relativistic electrons leaked from the X-ray PWN region and trapped at the forward bow shock. The other model, favored by \cite{2016ApJ...831..129R}, was the spectrum emitted by the ISM matter heated and compressed in a collisionless shock ({\tt SHELLS} model in \citealt{2013SSRv..178..599B}); it was consistent with the measured bow shock fluxes at reasonable values of the shock (pulsar) velocity and upstream ISM density.  

The bow shock of J2124 is dimmer and has a lower $S/N$, which prevents us from performing a similarly detailed analysis. Nevertheless, we can compare general properties of the J2124 shock with those of the J0437 shock. The FUV bow shocks of both pulsars are spatially coincident with the H$\alpha$ bow shocks (see Figure~\ref{hst}). In contrast to J0437, the J2124 bow shock shows a significant asymmetry of its shape with respect to the proper motion direction, which has been attributed to a local ISM density gradient perpendicular to the pulsar's direction of motion \citep{2002ApJ...580L.137G}. Interestingly, there is also an asymmetry in shock brightness, opposite in the the H$\alpha$ and FUV images (compare the top-left and bottom-right panels in Figure \ref{hst}); the reason of this asymmetry remains unclear. In addition to the asymmetry, the head (apex region) of the  FUV-H$\alpha$ J2124 bow shock is ``flatter'' than the head of the J0437 bow shock. This could be caused by anisotropy of the pulsar wind, such that the wind is concentrated in the pulsar's equatorial plane that is nearly perpendicular to the
pulsar's velocity \citep{2002ApJ...580L.137G,2014ApJ...784..154B}.

The FUV luminosity, $L(1250$--$2000\,{\rm \AA}) \approx 1.9\times 10^{29} d_{410}^2$ erg s$^{-1}$, extracted from the $A_s=52$ arcsec$^2$ area of the J2124 bow shock is a factor of $\approx 3.8d_{410}^2$ higher than the FUV luminosity from the $A_s=32$ arcsec$^2$ area of the J0437 shock. However, the (distance-independent) mean specific intensity, $I=L/(4\pi d^2 A_s) \approx 1.9\times 10^{-16}$ erg cm$^{-2}$ s$^{-1}$ arcsec$^{-2}$ for the J2124 bow shock, is a factor of 2.8 lower than that of the J0437 shock. Such a difference could be partly due to the fact that the width of the extraction region of the J2124 shock was somewhat broader than the actual width of the bright shock area. In addition, the more distant J2124 shock was imaged at larger physical separations in units of length from the apex, where the shock becomes intrinsically dimmer. Therefore, it is likely that the specific intensities of the J2124 and J0437 shocks are close to each other  at the same (physical) separations from the apices. Since in the ISM shock interpretation the emitted shock flux per unit emitting area is crudely proportional to the upstream density and only weakly depends on the pulsar velocity in the relevant range of parameters (see Figure 6 of \citealt{2016ApJ...831..129R}), it means that the ISM densities ahead of the J0437 and J2124 shocks
are close to each other.

For pulsar J0437, \cite{2016ApJ...831..129R} used the ratio of the F125LP and F140LP count rates, $C_{\rm F125}/C_{\rm F140} = 1.63\pm 0.08$, to estimate the slope of the J0437 spectrum in the FUV range, $-3\lesssim\alpha\lesssim -1$ (see Figure~4 in \citealt{2016ApJ...831..129R}). Similarly, for J2124, we have $C_{\rm F125}/C_{\rm F140} = 1.3\pm0.2$, from which we infer $-10\lesssim\alpha\lesssim -4$, i.e., a very steeply decreasing $f_\nu$ spectrum. This result, however, suffers from the large uncertainty of the F140LP count rate. Such a steep spectrum cannot be extrapolated to the optical (F475X) band because the extrapolated spectral flux would greatly exceed the upper limit in that filter, which requires a FUV-optical slope $\alpha\gtrsim 0$ (see Figure \ref{limits}), corresponding to $C_{\rm F125}/C_{\rm F140}\gtrsim 1.85$. Unless the count rate uncertainties are greatly underestimated due to unaccounted systematic errors, this discrepancy can be considered as an argument for a shock SED more 
complex than a simple PL.

In the shocked ISM emission model the ratio of FUV and H$\alpha$ fluxes depends on the model parameters. \citet{2014ApJ...784..154B} report the H$\alpha$ photon flux $\mathcal{F}_{\rm H\alpha}=5.3\times10^{-3}$\,photons\,s$^{-1}$\,cm$^{-2}$, or energy flux $F_{\rm H\alpha} = 1.6\times10^{-15}$\,erg\,s$^{-1}$\,cm$^{-2}$. This corresponds to the  ratio of F125LP to H$\alpha$ energy fluxes of $\approx6$, although this value can be larger (by a factor of up to 2) because we do not know the exact apex areas used by \citet{2014ApJ...784..154B}. For comparison, the flux ratio  in the same filters for J0437 is $F_{\rm FUV}/F_{\rm H\alpha}=12\pm 4$ (Figure~6 in \citealt{2016ApJ...831..129R}), i.e., generally consistent with
J2124. The non-detection of the J2124 bow shock in the F475X image provides additional constraints on the spectral shape and, possibly, on the emission mechanism. Figure~\ref{limits} shows a single PL model and a shocked ISM model (normalized to FUV flux measurement) compatible with the $3\sigma$ F475X upper limit. The shocked ISM model appears to be able to describe the measurements;
 however, given the complexity of the model and the scarcity of the data, we do not attempt to fit it. 

The detection of the FUV bow shock around two pulsars allows one to assume that such shocks can be detected from many other pulsars, including those from which no H$\alpha$ shock is seen because of a lack of neutral hydrogen atoms upstream of the shock. Such observations, currently possible only with the {\sl HST}, would be very useful for studying the ISM properties and the physics of relativistic shocks.

\acknowledgements
Support for program GO 13783 was provided by NASA through grant from the Space Telescope Science Institute, which is operated by by the Association of Universities for Research in Astronomy, Inc., under NASA contract NAS 5-26555. The authors acknowledge support from FONDECYT Regular Project 1150411, FONDECYT Postdoctoral Project 3150428, Center for Astronomy and Associated Technologies (CATA; PFB-06). We thank Denis Gonz\'alez-Caniulef for valuable help with the preparation of the observing proposal.

\facility{HST(ACS), HST(WFC3)}

\end{document}